\documentclass[conference]{IEEEtran}
\makeatletter
\def\ps@headings{%
\def\@oddhead{\mbox{}\scriptsize\rightmark \hfil \thepage}%
\def\@evenhead{\scriptsize\thepage \hfil \leftmark\mbox{}}%
\def\@oddfoot{}%
\def\@evenfoot{}}
\makeatother
\usepackage{amssymb,amsmath,amsthm}
\usepackage{graphicx}
 \usepackage{subfig}
\usepackage{mdwmath}
\usepackage{mdwtab}
\usepackage{soul}
\usepackage{cite}
 \usepackage{multirow}
 \usepackage{color} 
  \usepackage[linesnumbered,ruled]{algorithm2e}
\usepackage{setspace}
 
 \newtheorem{prop}{Proposition}
 \renewcommand{\vec}[1]{\mathbf{#1}}
\begin{document}
    \title{A Novel Network Coded Parallel Transmission Framework for High-Speed Ethernet  }
 \author{\IEEEauthorblockN{Xiaomin Chen$^*$, Admela Jukan$^*$ and Muriel M\'{e}dard$^{**}$}
\IEEEauthorblockA{Technische Universit\"at Carolo-Wilhelmina zu Braunschweig, Germany$^*$\\
Massachusetts Institute of Technology, US$^{**}$ } 
}
  \maketitle

\begin{abstract}   
 Parallel transmission, as  defined in high-speed Ethernet standards, enables to use  less expensive   optoelectronics  and offers   
backwards compatibility with legacy Optical Transport Network (OTN) infrastructure.  However, optimal parallel transmission does not scale to large networks, as it requires computationally expensive multipath routing algorithms to minimize differential delay, and thus the required buffer size, optimize  traffic splitting ratio, and ensure frame synchronization. In this paper, we propose a novel framework for high-speed Ethernet, which we refer to as \emph{network coded parallel transmission}, capable of effective buffer management and frame synchronization without the need for complex multipath algorithms in the OTN layer.  We show that using network coding can reduce the delay caused by packet re-ordering at the receiver, thus requiring a smaller overall buffer size, while improving the network throughput.  We design the framework in full compliance with high-speed Ethernet standards specified in IEEE802.3ba and present solutions for network encoding, data structure of coded parallel transmission, buffer management and decoding at the receiver side.  The proposed network coded parallel transmission framework is simple to implement and represents a potential major breakthrough in the system design of future high-speed Ethernet. 
 \end{abstract}
\section{Introduction}\label{intro}    
\par
\begin{figure} [ht]
  \centerline{\includegraphics[width=1\columnwidth]{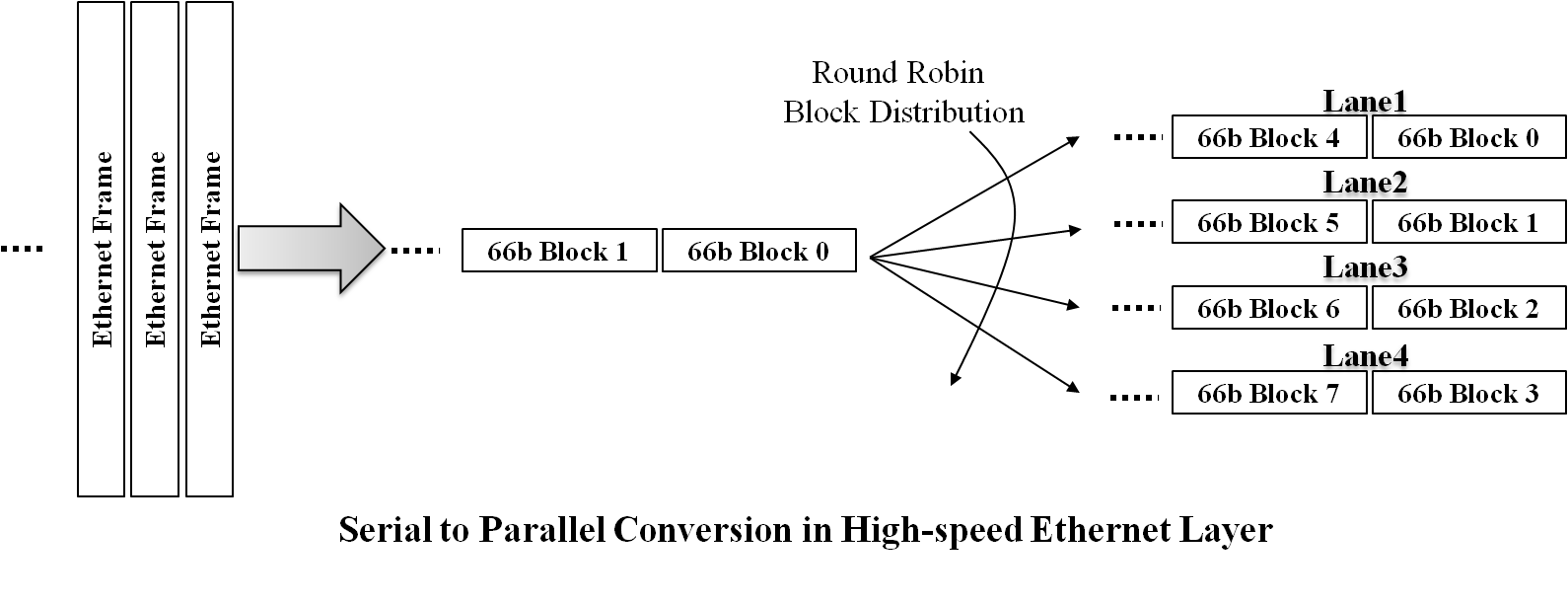}}
  \caption{Serial to parallel conversion of high-speed Ethernet traffic according to IEEE802.3ba~\cite{802.3ba}} 
  \label{Ethernetlayer}
      \vspace{-5 mm}
  \end{figure}

 IEEE 802.3ba standardizes high-speed Ethernet at transmission rates of 100Gbps and 40Gbps (100GE/40GE), with parallel transmission as one of the possible solutions.  Parallelism enables Ethernet to utilize slower but less expensive optoelectronic interfaces while achieving  high overall transmission speed. In addition, since most today's commercial optical networks are designed to support  transmission rates of 10Gbps or 40Gbps, parallelization enables networks to easily scale up to 100Gbps, and even beyond what is currently possible with the serial transmission. Figure \ref{Ethernetlayer} illustrates a 40GE parallel transmission system, with serial to parallel conversion of the Ethernet traffic. According to IEEE 802.3ba, instead of serially transmitting Ethernet frames at 40Gps, high speed Ethernet can utilize four parallel \emph{virtual Ethernet lanes} with lower data rates, i.e., 10Gps. Here, native Ethernet frames are first scrambled and regrouped into data blocks with 64 bits per block. Each block is then inserted two additional bits synchronization header, corresponding to either the data or control blocks, resulting to all blocks with same size, i.e., 66b. In a 40GE system, the 66b blocks are distributed to four Ethernet lanes in a round robin fashion.
 
  \begin{figure*} [ht]
  \centerline{\includegraphics[width=1.7\columnwidth]{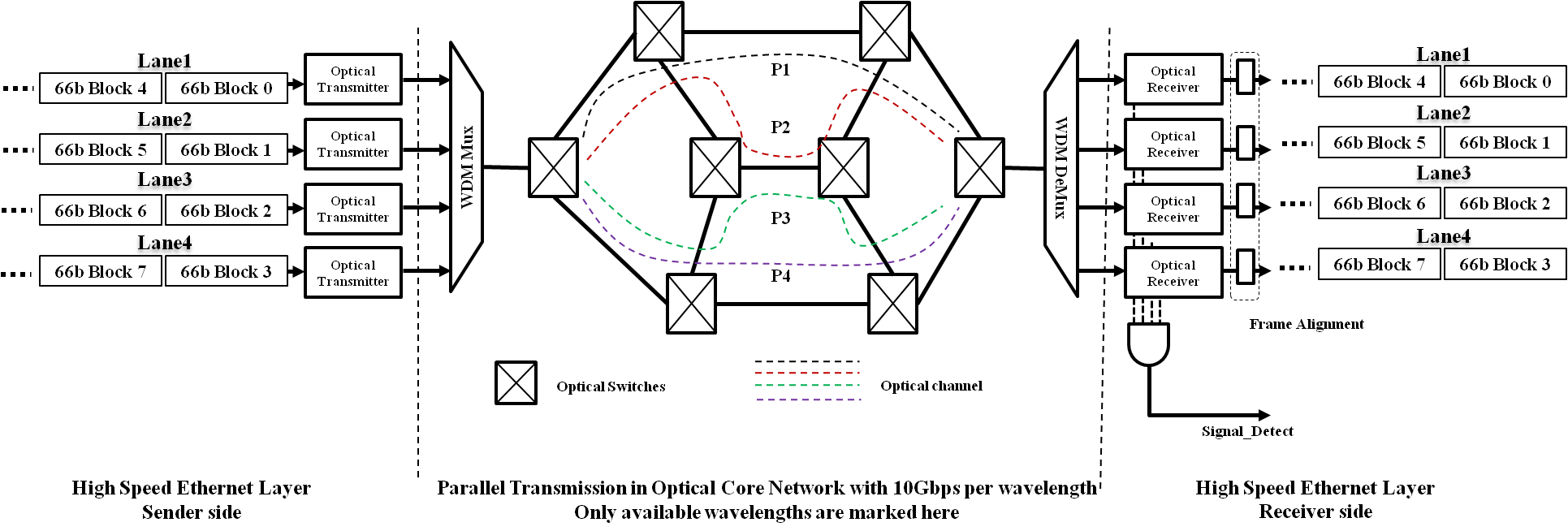}}
  \caption{Parallel transmission of 40Gbps Ethernet} 
  \label{Ethernet}
        \vspace{-4 mm}
  \end{figure*}
 
\par The dominance of the high-speed Ethernet technologies in the Local Area Networks (LAN) has now paved the way for their migration to metro and core networks. To this end, the parallel transmission across core networks requires \emph{routing}, i.e., single path or multipath. With single path routing, a path needs to be found with sufficient bandwidth to support multiple Ethernet lanes. In a Wavelength Division Multiplexing (WDM) network, for instance, multiple wavelengths would be allocated along the same fiber path. When single path routing is not possible, multipath routing can be used; Figure \ref{Ethernet} shows an example. Here, instead of searching for four wavelengths over the same fiber path, another set of wavelengths can be found along different paths. However, care needs to be taken of the so-called \emph{differential delay issue} caused by the paths' diversity. In our example, the traffic routed on path $P_1$ and $P_4$ (shorter paths) needs to be buffered until the traffic routed on $P_2$ and $P_3$ (longer paths) arrives to properly reorder the data blocks. Thus, to minimize the differential delay,  as well as to address issues of timing and synchronization, multipath routing algorithms need to be \emph{optimized},  which is computationally expensive. Moreover, it requires complex control implementations in the Optical Transport Network (OTN) layer, and does not scale for large networks \cite{Chen:JOCN:2012}.

\par In this paper, we propose a novel framework for high-speed Ethernet, which we refer to as network coded parallel  transmission. By applying network coding, we show that the requirements on optimality of multipath routing can be relaxed, optical control layer is kept simple and scalable,  while care can be taken of synchronization via network coding function at the end-systems.  Note that our scheme is indeed a network coding scheme, even if it is implemented at the source, because the code can be generated at different nodes and the type of codes we use, unlike traditional structured codes, be they block codes or rateless codes, are composable within the network without the need for decoding at intermiediate nodes. This characteristic is particulrly attractive for multipath systems, since no intermediate node may have sufficient degrees of freedom to decode. Even though generally re-encoding can occur at any node in the network, which could further improve the performance and completely eliminate the need for an routing algorithm, we show that a simple multipath routing algorithm is sufficient between the sender and receiver with network coding capability.  Using network coding can reduce the delay caused by re-ordering on the receiver side, thus requiring a comparably smaller buffer size, while improving the network throughput by timely realizing the paths carrying redundant coding information.  At the same time, our framework is in full compliance with high-speed Ethernet standard IEEE802.3ba, as it is presented by their detailed schemes for network encoding at the sender, data structure of coded parallel transmission, the buffer model and decoding at the receiver.  We show that by simply using a packet identifier we can ensure synchronization, and thus completely eliminate the need for synchronization through the OTN layer. The proposed network coded parallel transmission framework is simple to implement, and can be used as a promising solution in support of high-speed Ethernet.

\par The rest of the paper is organized as follows. In Section \ref{relatedwork}, we provide a brief literature review and summarize our contribution.   In Section \ref{problem}, we present the proposed network coded parallel transmission framework.    Finally, we show simulation results in Section   \ref{performance} and conclude the paper in  Section \ref{conclusion}.

\section{Related Work and Our Contribution}\label{relatedwork}
\par Since the early work from over a decade ago, the network coding has gained significant attention. \cite{Ahlswede:2000} showed that a multicast connection can achieve maximum flow capacity between source and any receiver using network coding, which is otherwise not achievable with traditional store-and-forward networking. Other than single source multicast network coding problem, multi-source multicast problem using network coding has also been studied. Here, multiple disjoint subgraphs are created of the given network, and network coding is applied simultaneously to multiple sessions.   \cite{Lun:2008}  proposed a linear programming based optimization model to find an optimal solution to construct the subgraphs. In \cite{Li:2003}, the efficiency of various coding schemes was studied for multicast connections and the simplest linear network coding was shown to be sufficient to reach the optimum, i.e., the max-flow between source and every receiver. In \cite{Koetter:TON:2003}, it has been shown algebraically that network coding can be reduced to operations on matrices, which allows for the use of random linear network coding.

\par Previous work on network coding in wireline networks has focused on connection protection and the network throughput of multicast connections. Menendez and Gannet \cite{Menendez:OFC:2008} proposed to use photonic XOR devices for network coding and show its operation with cross-session coding of two multicast sessions with a shared link. Liu et.al. \cite{Liu:2012} presented a scheme to apply network coding in optical networks. Manley et.al. \cite{Manley:JOCN:2010} provided an comprehensive study on all-optical network coding, and showed the effectiveness of using network coding for dedicated protection in the optical layer. Network coded multipath routing has been applied for erasure correction~\cite{Maxemchuk:2007}, where the combined information from multiple paths are transferred on a few additional paths. The additional information can be used to recover the missing information by decoding. The difference between this work and the previous work on coded multpath routing is that the previous work considered specific structured codes, generally Reed-Solomon or other Vandermonde matrix variations, which significantly restrict design parameters. In particular, these structured codes are not composable, unlike the codes we consider. This work considers random codes, which can be generated or regenerated at different Ethernet nodes in the network. Secondly, traditional block codes have significant restrctions on the choice of coefficients, which may not be attractive  for the purposes of buffer management and frame synchronization we consider here.

\par Furthermore, no prior work addresses applications of network coding for frame synchronization, buffer management and network throughput for high-speed Ethernet, to the best of our knowledge. We studied optical parallel transmission to support high-speed Ethernet in \cite{Chen:JOCN:2012} for the first time, with an optimized parallel transmission based on OTN/WDM networks. Such frameworks usually rely on the OTN layer for synchronization and routing information management. Similarly to our work, most multipath routing proposals assume the existence of a complex off-line optimization tool for minimization of differential delay and buffer dimensioning. Such assumptions do not provide solutions that scale to larger networks, or can be feasible for on-line implementations.  Our contribution in this paper is to design a network coded parallel transmission solution, which can be applied to any-size network with and without OTN layer and without complex multipath optimization algorithms. Our goal is to show that the overhead of introducing the network coding in the system design is a small price to pay to simplify the optical network control layer, as well as improve the system performance.
 \section{Network Coded Parallel Transmission}\label{problem}
\subsection{Preliminaries}\label{pre}
\par   Network coding referrers to the technique which allows a network node to combine the received packets based on 
   simple operations and send out encoded packet to the next hop.  It  has been widely used to address the performance bottlenecks in a network, such as to improve the network throughput of multicast connections.   To ease the understanding of the proposed framework, we first  present a brief  introduction of the relevant network coding principles. 
   
\par A network is represented as a directed graph $G (V,E)$, with $V$ and $E$ as the sets of network nodes and links, respectively. For every node $v\in V$, let $In(v)$ denote the set of incoming edges to $v$ and $Out(v)$ denote the 
set of outgoing edges from $v$.  Let us assume node $v$ is enabled with network coding capability, and receives one 
symbol  from each incoming edge $e' \in In (v)$, denoted as $y(e')$. The node $v$ sends out a encoded symbol on each outgoing edge $e \in Out(v)$, denoted as $y(e)$,  which is a linear combination of received symbols, i.e.,  $y(e) = \sum_{e'\in In(v)} m(e')y(e')$ where the coefficient   $m(e')$ is an element of \emph{local encoding vectors} which is randomly chosen in a certain finite field \textbf{F}. Note that  every arithmetic operation in network coding is over a certain finite field \textbf{F}. 

\par  When $v$ is the sender, i.e., $v=s$, a set of imaginary incoming edges $e'_1, e'_2,...,e'_\omega$   are introduced for 
generality. The number of imaginary channels are context dependent, which is denoted as $\omega$ here. Hence, the 
symbols sent to each link originating from sender, i.e., $e\in Out(s)$ are:   
$ y(e) = \sum_{e'\in In(s)} g(e')x(e')$
  Where  $\vec{g(e\rq{})}={g(e\rq{}_1),...,g(e\rq{}_\omega)}$ is given as  $\omega$-dimensional \emph{global encoding vector} and  $x_1, x_2,...,x_\omega$ are original symbols on $\omega$ imaginary incoming edges. 
   The relation between \emph{Global encoding vector} and \emph{local encoding vector} is given by
$g(e) = \sum_{e'\in In(v)} m(e')g(e')$. The decoding at any receiver is performed by Gaussian elimination on the received symbols of a given generation. For more details, please refer to \cite{Yeung:2006}.

Network coding introduces extra computational overhead in network nodes. Hence,  the encoding/decoding should 
be fast enough in practice to fully explore the advantages  of this technique. The performance of network coding 
operations strongly depends on the used coding method. Linear network coding has been widely applied and studied 
due to its simplicity. It has also been proven that  simple linear network coding is sufficient to achieve the maximum capacity of multicast connections in a lossless network with affordable computational overhead~\cite{Koetter:TON:2003}.

   \subsection{Reference Network Model}
\begin{figure} 
  \centerline{\includegraphics[width=0.999\columnwidth]{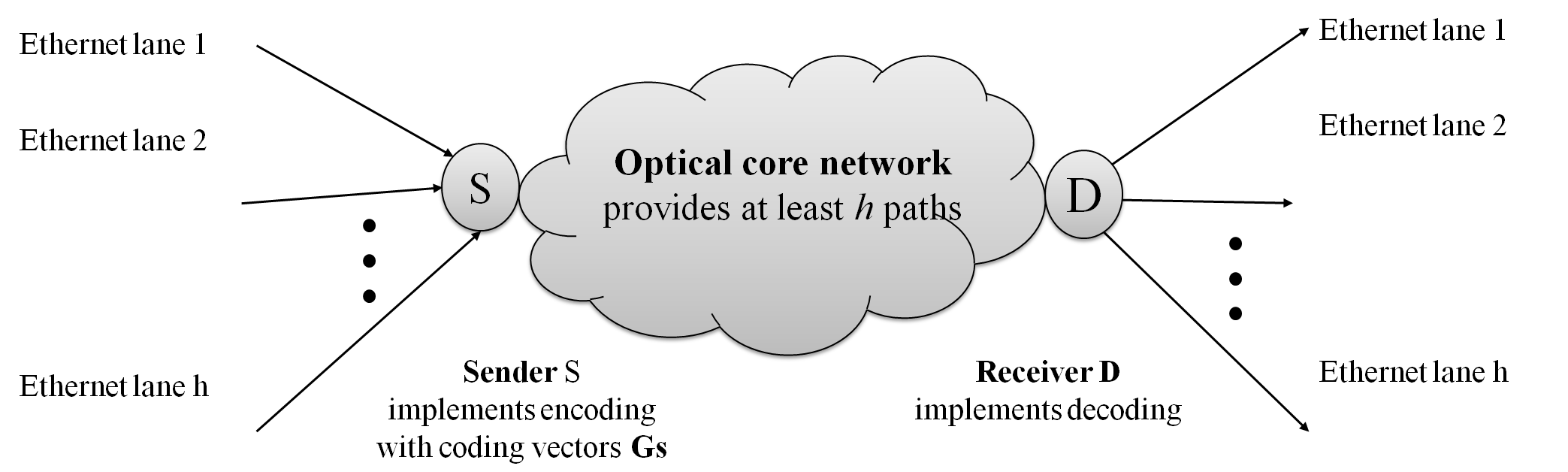}}
  \caption{Reference network model} 
  \label{fig:arch}
        \vspace{-5 mm}
  \end{figure}

   As already mentioned, network coding requires a network node to perform linear operations, such as 
   multiplication and addition over a certain field \textbf{F}.    To perform coding operations, it requires a local memory to temporarily store the received symbols to perform linear combinations. 
       However, this principle is not particularly practical for high-speed (optical)  networks since it would require the high-speed bit streams to "stop"
   for coding operations at every code.   In our framework, encoding and decoding therefore only happen in the end-systems, i.e., high-speed 
  Ethernet layer, while optical nodes in the core networks are treated as relay nodes that forward the coded 
  packets along the pre-defined paths. Although this requires some aspects of network control and routing, it is still simple and practical, compared to the  fast processing at every optical node. The reference network model of the proposed network coded parallel transmission system is shown in Figure \ref{fig:arch}. According to this model, if we assume there are $h$ Ethernet lanes,  the source node $S$ encodes the data blocks on each lane and sends out the packets to 
      multiple paths; these multiple paths are found with a simple and a fast online heuristic algorithm without typical constrains on differential delay, etc.
        At the destination $D$  the receiver buffers the received data and checks if it increases the rank of the received data matrix, i.e., to decide 
     whether the data is innovative. The decoding is successful when the received data matrix has a rank $h$.  Therefore,  the optical core networks have to be able to provide at least $h$ paths for the high-speed Ethernet, in order to receive sufficient  encoded data for decoding at the receiver side.

  \begin{prop} 
Network coded parallel transmission can be applied  only if the capacity of $min\_cut\{S\_D\} \geq h$ in the optical core network;  $S$, $D$ and $h$ are sender, receiver and number of Ethernet lanes, respectively.
\end{prop}
\begin{proof}
In order to recover the native symbols from the encoded symbol in destination $D$, a global vector $G_D$ is required has with a rank $h$, i.e., at least $h$ innovative symbols have to be received in $D$.  If the capacity of $min\_cut\{S\_D\}$ is less than $h$, then it is not possible to deliver $h$ innovative symbols between source and destination.    
\end{proof}
 
   \subsection{Network Coding in High-speed Ethernet Layer}
   Unlike the conventional parallel transmission where data blocks are sent to each lane in round robin fashion, the network coded parallel transmission takes data 
   blocks from all Ethernet lanes in parallel and sends out encoded data blocks to all outgoing links\footnote{As per nature of linear network coding, the size of the encoded data packet from all lanes is the same as the size of the original data packet, which makes this method extremely efficient.}. The virtual Ethernet lanes are modeled as the incoming edge of the source 
  node, i.e.,  $In(s)={e\rq{}_1, e\rq{}_2,..., e\rq{}_h}$, where $h$ is the number of Ethernet lanes. For instance, $h=4$ in case of 40GE standard~\cite{802.3ba}.

     \begin{figure} 
  \centerline{\includegraphics[width=1\columnwidth]{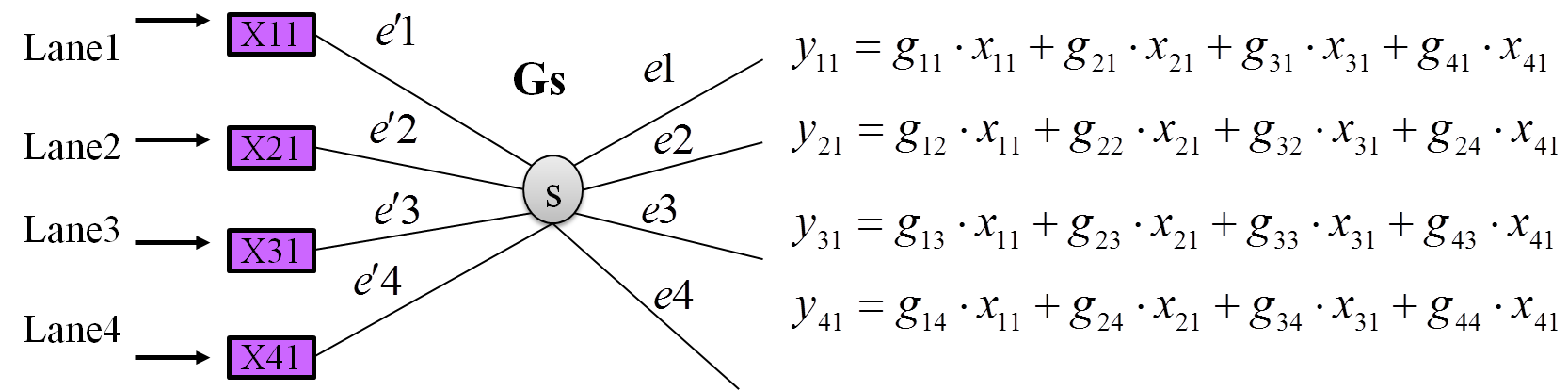}}
  \caption{Encoding over parallel lanes of an 40Gbps Ethernet} 
  \label{fig:code}
        \vspace{-5 mm}
  \end{figure}
   
\par    In the network coded framework, a basic unit for coding is referred to as \emph{symbol} and the size of a symbol is context dependent, which is determined by the finite field \textbf{F}. For example, a symbol 
   contains 8 bits when the size of field \textbf{F} is $GF(2^8)$.  Let us denote a \emph{packet} as $\vec{x}_i$, which is composed of $N$ symbols. All the symbols in a packet are encoded at the same time.  Assume a symbol in packet $\vec{x}_i$  is denoted as $x_{ij}$, where $i$ is the index of  the Ethernet lane, $i=1,2..,h$ and $j$ is the index of the symbols in this packet, $j= 1,2,...,N$, the encoding in the source node is defined as follows, i.e., 
    \begin{equation}\label{source}
 \begin{pmatrix}
  \vec{y}_1\\
 \vec{y}_2\\
  \vdots   \\
\vec{y}_h
 \end{pmatrix}
 =
 \begin{pmatrix}
  y_{11}&y_{12}&\cdots&y_{1N}\\
 y_{21}&y_{22} &\cdots&y_{2N}\\
  \vdots&\vdots&\ddots&\vdots   \\
  y_{h1} &y_{h2}&\cdots&y_{hN}
  \end{pmatrix}
 =
   \vec{G}_s
 \cdot
   \begin{pmatrix}
\vec{x}_1\\
 \vec{x}_2\\
  \vdots  \\
   \vec{x}_h 
  \end{pmatrix}
   \end{equation}
   where $\vec{y}_i$ is the encoded packet on lane $i$ and $\vec{G}_s$ is the coefficient matrix at source node $s$ for encoding. The coefficient matrix at source node $\vec{G}$ and the data matrix before encoding are defined as
    $\vec{G}_s=
    \begin{pmatrix}
g_{11}     & g_{21}    &\cdots & g_{h1} \\
g_{12}    & g_{22}  &\cdots & g_{h2}  \\
  \vdots       & \vdots  & \ddots     & \vdots  \\
  g_{1h} & g_{2h}  &\cdots & g_{hh}
 \end{pmatrix}  
      $ and  
$ \begin{pmatrix}
\vec{x}_1\\
 \vec{x}_2\\
  \vdots  \\
   \vec{x}_h 
  \end{pmatrix}
  =
  \begin{pmatrix}
  x_{11}&x_{12}&\cdots&x_{1N}\\
 x_{21}&x_{22} &\cdots&x_{2N}\\
  \vdots&\vdots&\ddots&\vdots   \\
  x_{h1} &x_{h2}&\cdots&x_{hN}
  \end{pmatrix}
  $, respectively.

\par For easier understanding, we present an example of encoding for 40GE where $h=4$ as shown in Figure \ref{fig:code}. In this example, one symbol is shown on each lane, i.e., $x_{11}$ on lane 1, $x_{21}$ on lane 2 and $x_{31}$ and $x_{41}$ on lane 3 and lane 4, respectively. The encoded symbols are the product of encoding vectors and original symbols. For instance, the first encoded symbol on path $p_1$ ($e_1$ in Figure \ref{fig:code}) is calculated as $y_{11}=g_{11} \cdot x_{11}+ g_{21} \cdot x_{21}+g_{31} \cdot x_{31}+g_{41} \cdot x_{41}$. The complete data structure of encoding over parallel lanes for 40GE is shown in Figure \ref{Ethernet2}.  To encoding over 4 parallel Ethernet lanes over field $GF(2^8)$, a  $4\times4$ encoding coefficient matrix is required, with each encoding coefficient is randomly chosen over field $GF(2^8)$. As previously mention, it is important to note that the encoded symbols have the same length as the original symbols, for instance, $y_{11}$ has the same length of $x_{11}$. This makes the whole concept extremely efficient to implement.

\subsection{Packet Format and Generation Partitioning}\label{packet}
\begin{figure} [ht]
  \centerline{\includegraphics[width=1\columnwidth]{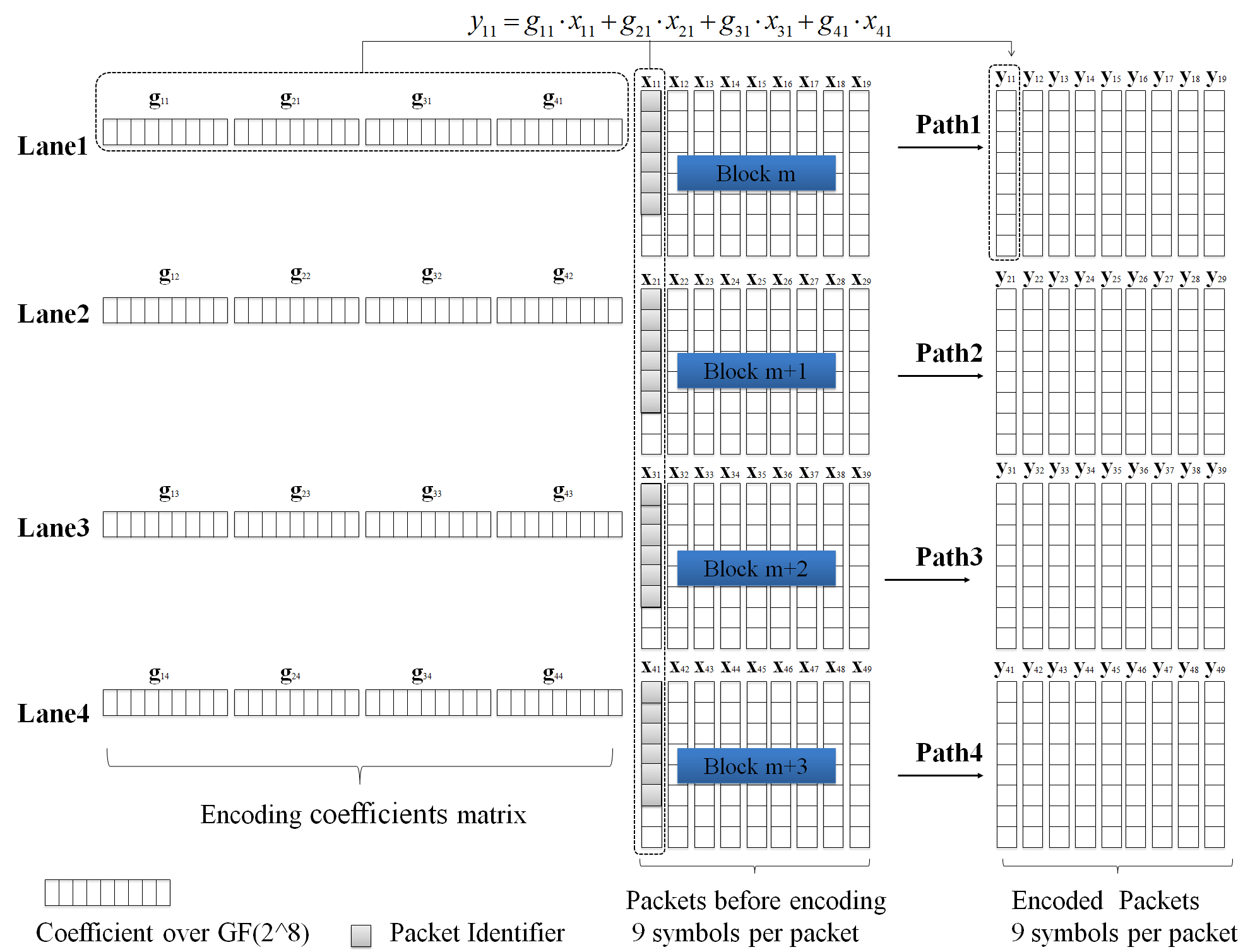}}
  \caption{Data structure of the 40GE example with encoding} 
  \label{Ethernet2}
  \vspace{-3 mm}
  \end{figure}

As specified in IEEE 802.3ba, high-speed Ethernet is standardized to use 64b/66b \emph{encoding}\footnote{Note 
that the term \emph{encoding} here is different from network coding. It refers  to   a line code that transforms 64-bit 
data to 66-bit line code to provide enough state changes to allow reasonable clock recovery and facilitate alignment of 
the data stream at the receiver.} in the Physical Coding Sublayer (PCS) \cite{802.3ba}. Two extra bits added to the 64 bit payload are used as
the synchronization header, with $``01"$ denoting the block carries data while $``10"$ denoting the block carriers control information. Let us assume that the encoding coefficients are chosen over the field $GF(2^8)$, i.e., a symbol contains 8 bits.
A packet shown in Figure \ref{Ethernet2} contains of 9 symbols, which is in fact a 66b 
data block with   with 6 bits used as the packet identifier.  However, the size of a packet 
is flexible in different systems. For instance, an Ethernet frame with 1500 bytes can be formulated as  a packet for network encoding, which contains 1500 symbols of field size is $GF(2^8)$ or 750 symbols if the field size is $GF(2^{16})$.  

Figure \ref{Ethernet2} shows the data structure of encoding over 4 lanes for parallel transmission of 40GE. A packet in 
this example is $\vec{x}_i=\{x_{i1}, x_{i2},...,x_{i9}\}$, and $i= 1,2,3,4$. All the packets that are related to the same 
set of encoding vectors are referred to as in the same  \emph{generation}, which are required to be decoded at the 
same time.  In the example shown in Figure \ref{Ethernet2}, the data blocks, i.e., block $m$, block $m+1$, block 
$m+2$ and block $m+3$ are in the same generation. The receiver can successfully perform decoding only when it receives all four encoded packets along four paths, i.e., $\vec{y}_1$,  $\vec{y}_2$, $\vec{y}_3$ and $\vec{y}_4$.
The generation is  obtained by $\lfloor{PacketIdentifier/h}\rfloor$. The 6-bit field used for packet identifier is reset after every 64 blocks.

 \subsection{Buffer Early Release}
\par As previously mentioned, current approaches to multipath routing require optimizations. Due to packet reordering, a buffer is required in addition to cache the packets routed on the shorter paths until the packets routed on the longest path are received.  This is a challenge, especially in high speed networks, such as 100GE.  
Network coding can address this challenge, where unlike with the multipath routing, the packets can be released from the buffer in the actual process of decoding, thus requiring at least the same buffer size as for optimal multipath routing, or smaller. This, on the other hand has a direct implication on network throughput, which can be drastically improved. Refer back to the 40GE example shown in Figure \ref{Ethernet} for illustration of this feature this principle.  
    
 \par
\begin{figure*} [ht]
 \centerline{\includegraphics[width=1.75\columnwidth]{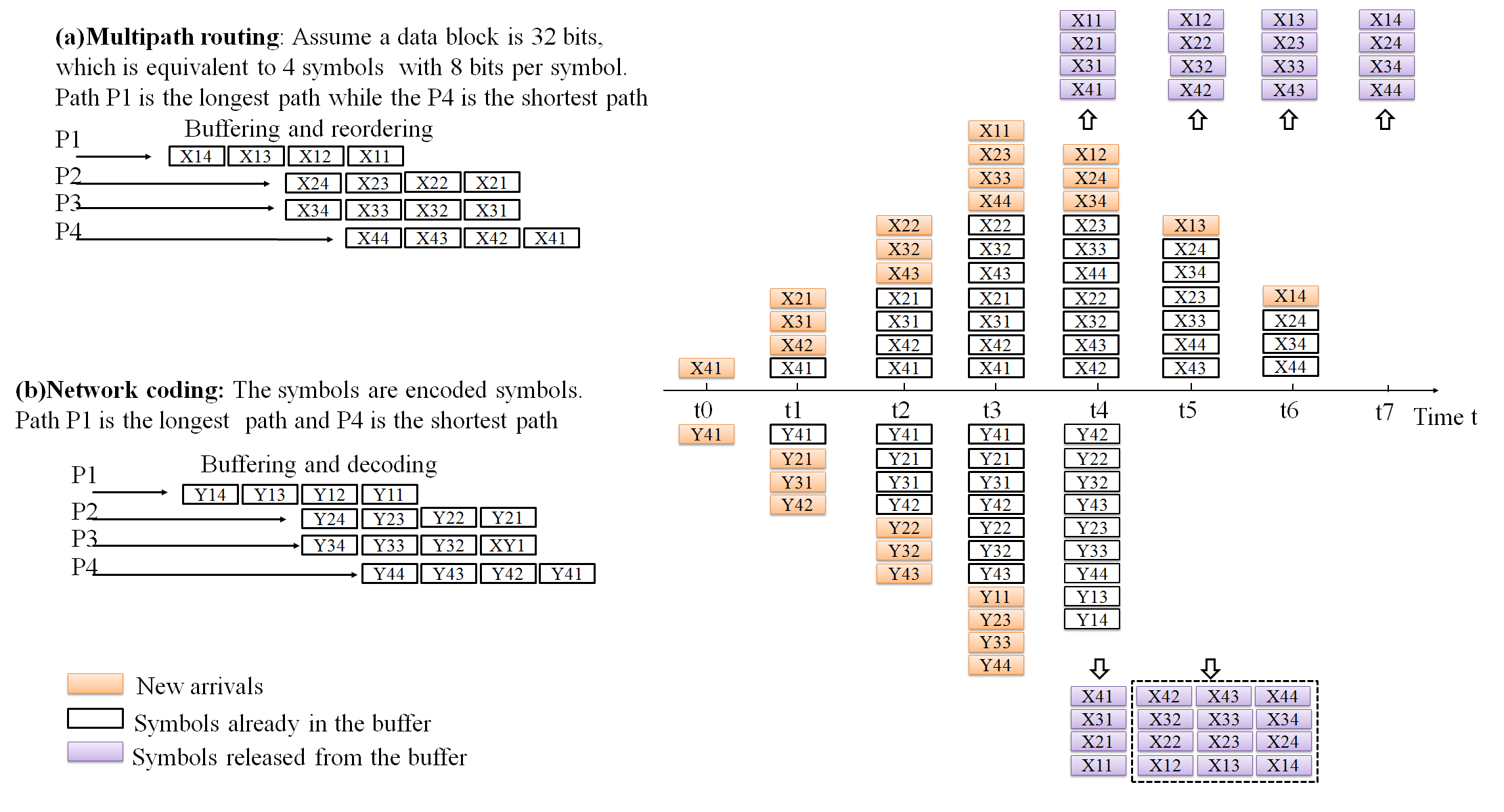}}
\caption{Buffer status at the receiver for multipath routing and network coded parallel transmission} 
\label{Comp}
      \vspace{-3 mm}
\end{figure*}

 Let us assume that $P_1$ has the largest end-to-end delay and $P_4$ is the shortest path, while $P_2$ and $P_3$ have the same end-to-end delay.  Figure \ref{Comp} shows the buffer status for multipath routing, and network coded parallel transmission, respectively. In this example, 16 symbols are sent, with 4 symbols on each path. In each time unit, the destination receives a symbol on each path. It can be seen that the receiver can start decoding at time $t_3$ upon the arrival of encoded packet $Y_{41}$ from $P_1$. As a result, $X_{11}, X_{21}, X_{31}$ and $X_{41}$ are decoded and released from the buffer. At the same time, the multipath routing method receives symbols $X_{11}, X_{23}, X_{33}$ and $X_{44}$ at time $t_3$ and has to perform the re-ordering process (via an algorithm) at time $t_4$. The re-ordered symbols are released at time $t_5$. Here, at time $t_4$, the buffer is already empty in the network coded scheme, while the multipath routing method needs to wait until $t_7$ to release all symbols from the buffer.  This example illustrates that the network coded parallel transmission can significantly alleviate the buffer size requirement.

 \section{Numerical Results}\label{performance} 
\par In our performance study, traffic is generated in the sender as a bit stream encapsulated into  66b data blocks, later grouped as packets for network coding.  Each packet is associated with a packet ID, while the packet size is flexible and can be defined by the end-system. For instance, a 66b data block has 6 bits added as the packet identifier.  One of the  performance measures we focus on is  \emph{packet loss ratio}, defined as the percentage of dropped packets due to the overflow of the buffer at the receiver. The buffer is implemented as a shared buffer which can be accessed by all paths while it outputs a single data stream. The reordering and decoding follow the same principle shown in Figure \ref{Comp}. The buffer size is defined in the number of packets $m$ that can be actually buffered. The values of $m$ under study are set to be 1000, 2000, 3000 and 4000 packets, respectively. For instance, when a packet is 9 bytes, buffer size is 9KB for $m=1000$. 
In each experiment,  approximately 20,000 packets are sent from sender to receiver and each experiment is repeated at least 100 times to obtain a stable average value. The field size used is $GF (2^8)$ in the results presented here, i.e., every \emph{symbol} contains 8 bits. The encoding matrix is randomly generated over field $GF (2^8)$. 

\par Here, we majorly focus on the results that compare the proposed network coded parallel transmission framework with the conventional multipath routing. Both the multipath and network coded parallel transmission schemes are implemented in Java and the simulation experiments are scaled down for feasibility. Without loss of generality, the sending rates under study are 800Kbps, 1.6Mbps and 4Mbps, respectively; thus the values obtained in the results can be linearly scaled up to the values in real systems, 800Mbps, 1.6Gbps and 4Gbps respectively.  Also without loss of generality, four paths are simulated and are assumed to have the same bandwidth, i.e., traffic is equally distributed to each path. For instance, when the sending rate is 4Mbps, each path has a line rate of 1Mbps. The differential delay between each path is arbitrarily defined and has not been calculated based on routing.

      \vspace{-2 mm}

   \begin{figure} [ht]
 \centerline{\includegraphics[width=0.95\columnwidth]{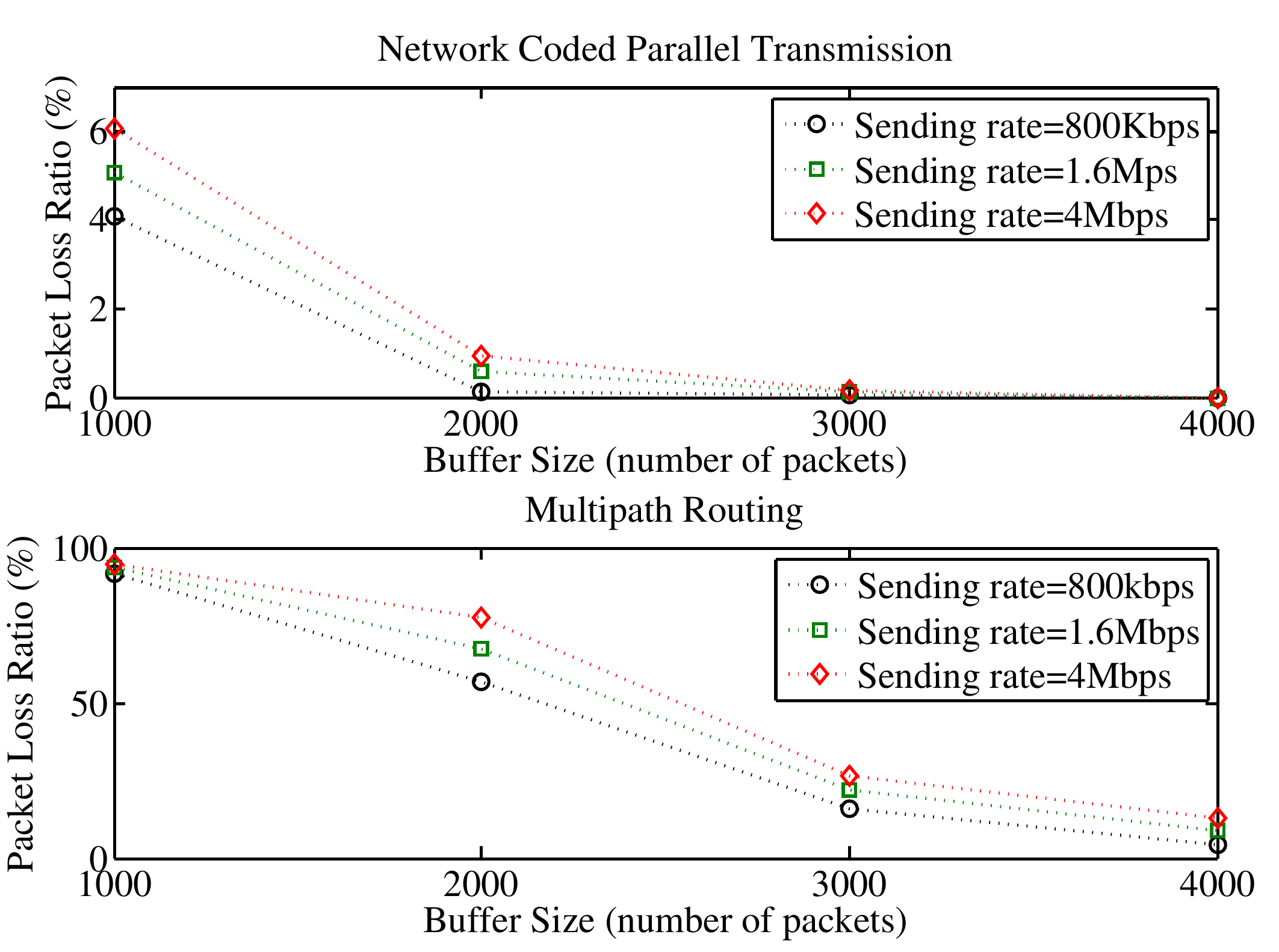}}
\caption{Packet loss ratio with different buffer sizes} 
\label{fig:buffer:100}
      \vspace{-4 mm}
\end{figure}
   \begin{figure} [ht]
 \centerline{\includegraphics[width=0.95\columnwidth]{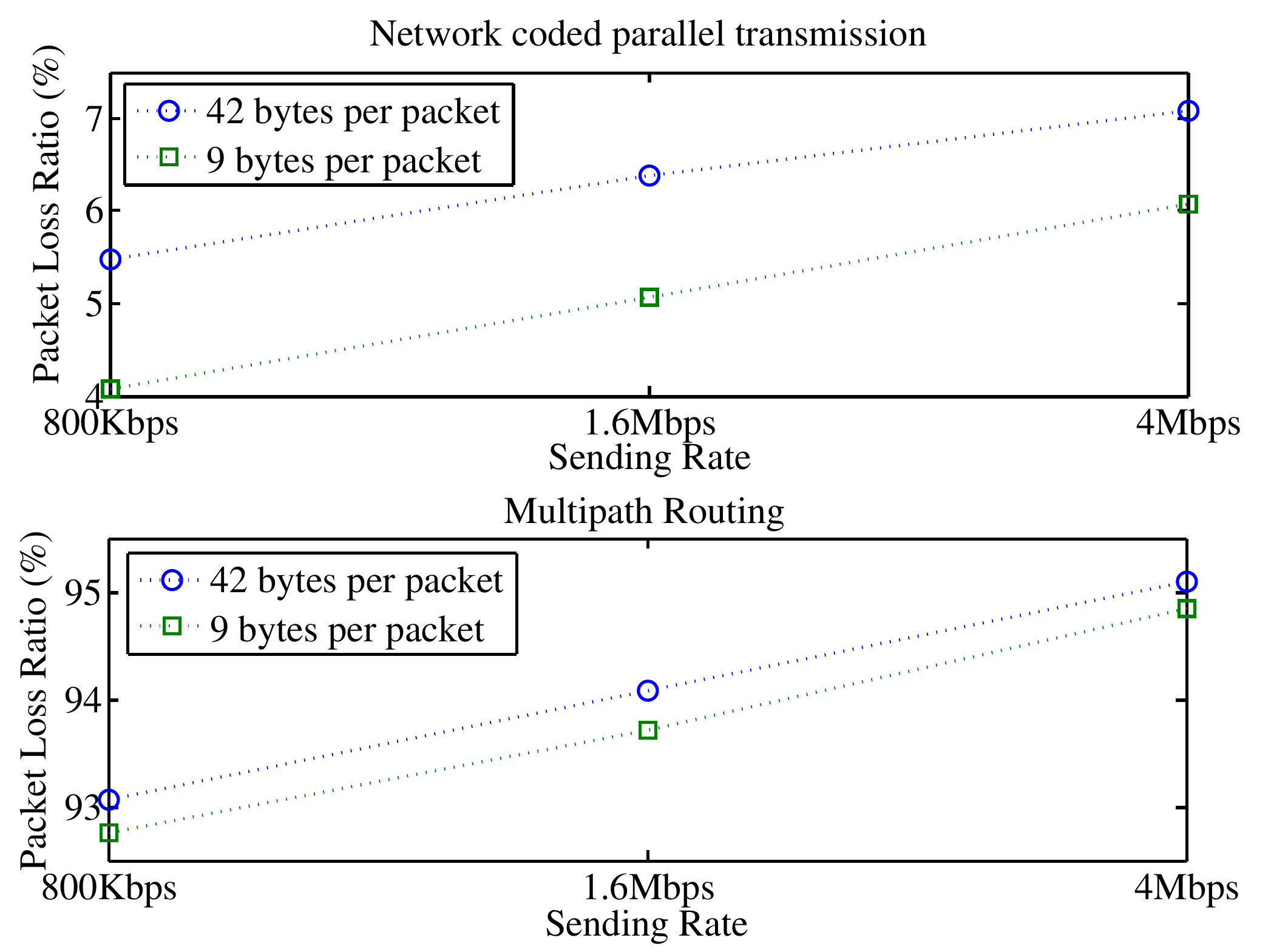}}
\caption{Packet loss ratio with different packet sizes} 
\label{fig:size}
      \vspace{-1 mm}
\end{figure}

\par Figure \ref{fig:buffer:100} shows the impact of buffer limit on both network coded and multipath routing schemes.  The delay of the four paths are   $300 ms$, $400 ms$, $500 ms$ and $600 ms$, respectively; and packet size is set to be 9 symbols. The differential delay between each path is sufficiently large (i.e., for the assumed system design values) to compare the performance of both schemes in terms of buffer release speed, which is directly proportional to the required buffer size, and also the resulting network throughput. In other words, the faster the buffer release speed, the smaller the required buffer size, and the smaller the amount of data transmitted over the network. It can be seen that applying network coding leads to a very small packet loss as compared to multipath routing. This is because multiple generations can be decoded at the same time, if sufficient number of innovative packets are in the buffer.  The different sending rates make for a small difference in the results with same buffer size, due to the fact that the differences between sending rates are not very significant. It can also be seen that the packet loss ratio decreases in both cases when buffer size increases. When buffer size is larger than 2000 packets, there is almost no packet loss with network coding, while multipath routing still shows a comparably high packet loss ratio,e.g, around $77\%$ when sending rate is 4Mbps.

\par It should be noted, however, that when the packet size increases, the  packet loss ratio in the network coded parallel transmission also increases, as it requires longer time to decode packets. At the same time, the multipath routing scheme uses re-ordering (packet storage and reordering process while waiting for all packets to arrive), and is thus not directly affected by the packet size.  As shown in Figure \ref{fig:size}, when a packet contains five 66b data block and 6 bit packet identifier, i.e., 42 bytes per packet, the packet loss ratio increases around $1\%-1.5\%$ in case of network coded parallel transmission. In spite of these results and the delay due to the longer decoding times, the network coded parallel transmission has much lower packet loss ratio compared to multipath routing.

 \section{Conclusion}\label{conclusion}
\par In this paper, we proposed a novel network coded parallel transmission framework for high-speed Ethernet in full compliance with IEEE802.3ba. We presented detailed schemes for network encoding at the sender, data structure of coded parallel transmission, the buffer model and decoding at the receiver.  The results are promising as they show that the requirements on optimality of multipath routing can be indeed relaxed by applying network coding. Also, the buffer required by parallel transmission is significantly smaller when the simple linear network coding is applied, which can have major implications on system design at high speeds, at 40Gb/s and plus. The proposed network coded parallel transmission framework is simple to implement, and carries significant practical potential. Our future work will consider the use of subgraphs beyond multipath schemes, in particular by incorporating coding in the interior of the network. Such generalizations should provide us considerable flexibility in route planning and load balancing, while maintaining very light state information.
 \bibliographystyle{IEEEtran}
\bibliography{codingbib}
\end{document}